\documentclass[a4paper, oneside, twocolumn, notitlepage, 10pt]{extarticle_ecoc}
\usepackage{gensymb}
\usepackage{ecoc}
\usepackage{pgfplots,amsmath}
\usepackage{multirow}
\usepackage{titlesec}
\titlespacing{\section}{0pt}{4pt plus 1pt minus 1pt}{1.5pt}

\begin{document}
\selectlanguage{english}    

\title{Single-Step Digital Backpropagation for O-band Coherent Transmission Systems}


\author{
    Romulo~Aparecido\textsuperscript{(1,$*$)}, 
    Jiaqian~Yang\textsuperscript{(1)},
    Ronit~Sohanpal\textsuperscript{(1)},
    Zelin~Gan\textsuperscript{(1)},
    Eric~Sillekens\textsuperscript{(1)},
    \\
    John~D.~Downie\textsuperscript{(2)},
    Lidia~Galdino\textsuperscript{(2)},
    Vitaly~Mikhailov\textsuperscript{(3)},
    Daniel~Elson\textsuperscript{(4)},
    Yuta~Wakayama\textsuperscript{(4)},
    \\
    David~DiGiovanni\textsuperscript{(3)},
    Jiawei~Luo\textsuperscript{(3)},
    Robert~I~Killey\textsuperscript{(1)}, and
    Polina~Bayvel\textsuperscript{(1)}
}

\maketitle                  


\begin{strip}
    \begin{author_descr}

         \textsuperscript{(1)} Optical Networks Group, UCL (University College London), London, UK. \\\textsuperscript{$*$}\textcolor{blue}{\uline{romulo.aparecido.22@ucl.ac.uk}} \\
        \textsuperscript{(2)} Corning Research and Development Corporation, Corning, NY 14831, USA\\
        \textsuperscript{(3)} Lightera Labs, Somerset, NJ 08873, USA\\
        \textsuperscript{(4)} KDDI Research, Inc., 2-1-15 Ohara, Fujimino 356-8502, Japan

    \end{author_descr}
\end{strip}

\renewcommand\footnotemark{}
\renewcommand\footnoterule{}


\begin{strip}
    \begin{ecoc_abstract}
        We demonstrate digital backpropagation-based compensation of fibre nonlinearities in the near-zero dispersion regime of the O-band. Single-step DBP effectively mitigates self-phase modulation, achieving SNR gains of up to 1.6~dB for 50~Gbaud PDM-256QAM transmission over a 2-span 151~km SMF-28 ULL fibre link.\textcopyright2025 The Author(s)
    \end{ecoc_abstract}
\end{strip}


\section{Introduction} \vspace{-0.3em}

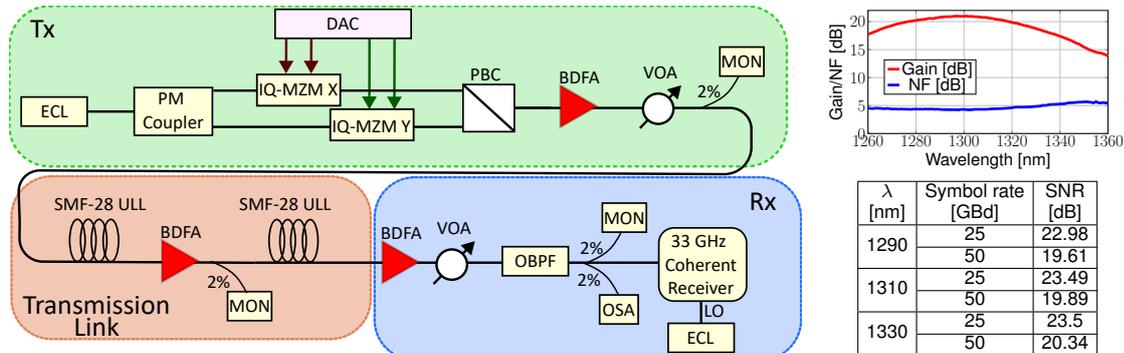
\begin{figure*}[b]
\vspace{-1em}
  \centering
  \begin{minipage}{0.62\textwidth}
    \centering
    \includegraphics[width=1.05\linewidth]{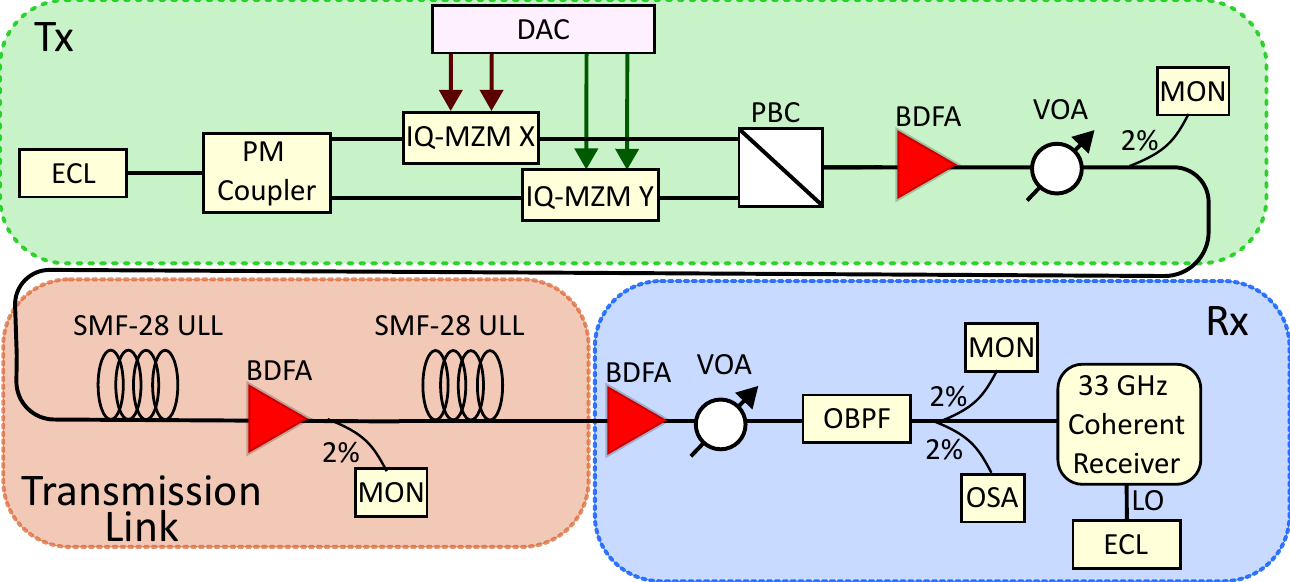}
  \end{minipage}%
  \hspace{-0.02\textwidth}  
  \begin{minipage}{0.38\textwidth}
    \centering
    \resizebox{0.65\linewidth}{!}{\begin{tikzpicture}
\begin{axis}[
    width=14cm,
    height=8cm,
    xlabel={\Huge Wavelength [nm]},
    ylabel={\Huge Gain/NF [dB]},
    ylabel style = { yshift=10pt},
    xlabel style = { yshift=-10pt},
    xmin=1260, xmax=1360,
    ymin=0, ymax=22,
    xtick={1260, 1280, 1300, 1320, 1340, 1360},
    axis y line*=left,
    axis x line*=bottom,
    axis lines=box,
    xticklabel style={
        /pgf/number format/.cd,
        use comma=false,
        1000 sep={}
    },
    tick label style={font=\Huge},
    legend style={font=\Huge, at={(0.45,0.6)}, anchor=north east},  
    grid=both,
    scaled x ticks=false
]

\addplot[color=red,  line width=4pt] coordinates {
(1260,17.7565) (1261,17.8905) (1262,18.0590) (1263,18.2314) (1264,18.3851)
(1265,18.6091) (1266,18.7403) (1267,18.9198) (1268,19.0501) (1269,19.1496)
(1270,19.3024) (1271,19.4360) (1272,19.5682) (1273,19.6478) (1274,19.7610)
(1275,19.8493) (1276,19.9438) (1277,20.0397) (1278,20.0653) (1279,20.1528)
(1280,20.2130) (1281,20.3134) (1282,20.3490) (1283,20.3805) (1284,20.4427)
(1285,20.5460) (1286,20.5907) (1287,20.5173) (1288,20.6466) (1289,20.7044)
(1290,20.7792) (1291,20.8256) (1292,20.8770) (1293,20.8659) (1294,20.8845)
(1295,20.9049) (1296,20.9649) (1297,20.9811) (1298,20.9585) (1299,20.9392)
(1300,20.9646) (1301,21.0025) (1302,20.9113) (1303,20.9451) (1304,20.8848)
(1305,20.8756) (1306,20.8437) (1307,20.7795) (1308,20.7458) (1309,20.7740)
(1310,20.6706) (1311,20.5978) (1312,20.5499) (1313,20.4444) (1314,20.3842)
(1315,20.3509) (1316,20.2468) (1317,20.1896) (1318,20.0561) (1319,19.9798)
(1320,19.8535) (1321,19.7582) (1322,19.7133) (1323,19.5891) (1324,19.4632)
(1325,19.3287) (1326,19.2447) (1327,19.1340) (1328,19.0175) (1329,18.8905)
(1330,18.7576) (1331,18.6162) (1332,18.5173) (1333,18.3261) (1334,18.2369)
(1335,18.0679) (1336,17.9329) (1337,17.8299) (1338,17.7101) (1339,17.5444)
(1340,17.3915) (1341,17.2360) (1342,17.0800) (1343,16.8882) (1344,16.7336)
(1345,16.5474) (1346,16.3273) (1347,16.1190) (1348,15.9998) (1349,15.6887)
(1350,15.4500) (1351,15.2558) (1352,15.0108) (1353,14.8558) (1354,14.7064)
(1355,14.5150) (1356,14.4567) (1357,14.3711) (1358,14.2470) (1359,14.0985)
(1360,13.8436)
};

\addplot[color=blue, line width=4pt] coordinates {
(1260,4.5418) (1261,4.4825) (1262,4.4571) (1263,4.5289) (1264,4.5158)
(1265,4.5035) (1266,4.4461) (1267,4.4514) (1268,4.4756) (1269,4.4465)
(1270,4.3683) (1271,4.3907) (1272,4.3744) (1273,4.3512) (1274,4.3834)
(1275,4.3537) (1276,4.3525) (1277,4.3334) (1278,4.3186) (1279,4.3705)
(1280,4.3575) (1281,4.3594) (1282,4.3624) (1283,4.3503) (1284,4.3922)
(1285,4.3517) (1286,4.3917) (1287,4.3526) (1288,4.3319) (1289,4.3535)
(1290,4.3109) (1291,4.2818) (1292,4.2636) (1293,4.2672) (1294,4.2483)
(1295,4.2531) (1296,4.2622) (1297,4.2545) (1298,4.2240) (1299,4.3237)
(1300,4.2999) (1301,4.2638) (1302,4.2377) (1303,4.2385) (1304,4.3166)
(1305,4.3240) (1306,4.3216) (1307,4.3348) (1308,4.3199) (1309,4.3326)
(1310,4.3391) (1311,4.3913) (1312,4.4196) (1313,4.4316) (1314,4.4289)
(1315,4.4574) (1316,4.4580) (1317,4.5114) (1318,4.5185) (1319,4.5620)
(1320,4.6065) (1321,4.6881) (1322,4.7183) (1323,4.7248) (1324,4.7066)
(1325,4.7173) (1326,4.7434) (1327,4.7533) (1328,4.7666) (1329,4.8070)
(1330,4.8361) (1331,4.8742) (1332,4.9250) (1333,5.0535) (1334,5.0098)
(1335,5.1264) (1336,5.1785) (1337,5.2108) (1338,5.2162) (1339,5.2496)
(1340,5.2753) (1341,5.2960) (1342,5.3194) (1343,5.4330) (1344,5.4919)
(1345,5.5099) (1346,5.5169) (1347,5.5946) (1348,5.5994) (1349,5.6187)
(1350,5.6401) (1351,5.6672) (1352,5.6225) (1353,5.5818) (1354,5.4849)
(1355,5.6928) (1356,5.5038) (1357,5.4537) (1358,5.5421) (1359,5.4511)
(1360,5.5094)
};

\legend{Gain [dB], NF [dB]}

\end{axis}
\end{tikzpicture}}
    
    \vspace{0.1cm}  

    \scriptsize
    \setlength{\tabcolsep}{3pt}  
    \renewcommand{\arraystretch}{1}  
    \begin{tabular}{|c|c|c|}
        \hline
        $\lambda$ & Symbol rate & SNR \\ 
        {[nm]} & {[GBd]} & {[dB]} \\ \hline
        \multirow{2}{*}{1290} & 25 & 22.98 \\ \cline{2-3} 
                          & 50 & 19.61 \\ \hline
        \multirow{2}{*}{1310} & 25 & 23.49 \\ \cline{2-3} 
                          & 50 & 19.89 \\ \hline
        \multirow{2}{*}{1330} & 25 & 23.5  \\ \cline{2-3} 
                          & 50 & 20.34 \\ \hline
      \end{tabular}
  \end{minipage}

  \caption{O-band experimental transmission setup with measured B2B SNR values. The inset shows the gain and noise figure (NF) as a function of the wavelength for an input of 0 dBm. External cavity laser (ECL), polarisation maintaining (PM), in-phase-quadrature Mach-Zehnder modulator (IQ-MZM), polarisation-beam combiner (PBC), bismuth-doped fibre amplifier (BDFA), variable optical attenuator (VOA), optical bandpass filter (OBPF), power monitor (MON), local oscillator (LO).}
  \label{fig:setup}
\end{figure*}

\noindent Optical networks are central to the global digital communications infrastructure~\cite{Winzer:18,bayvel2016maximizing} with cloud services and AI applications driving capacity demands of inter-/intra-data centre networks. To meet this growth, research has focused on accessing the available fibre bandwidth beyond C-band~\cite{2024JOpt}.  The O-band, centred at 1310~nm, offers a number of benefits: it has a large bandwidth (17.5 THz) relative to the C-band; the low dispersion of standard single-mode fibre across the O-band avoids the need for large dispersion compensating filters; and it is located, spectrally, well away from the C-band, avoiding interchannel stimulated Raman scattering (ISRS) between the bands in the case of O+C-band transmission~\cite{10568315}.
Recent developments in bismuth-doped fibre amplifiers (BDFAs) make it possible to use a single amplifier to cover the bandwidth of the O-band ~\cite{8696882, 9765488}, eliminating the need for multiple parallel amplifiers and associated WDM filters, used to split and combine the different bands.
However, O-band transmission also comes with some drawbacks: signals experience higher attenuation, approximately 0.3 dB/km, due to increased Rayleigh scattering in silica, whereas C-band signals typically experience attenuation of between 0.15 dB/km and 0.2 dB/km.
Additionally, the larger nonlinear coefficient and near-zero dispersion of standard single-mode fibre in the O-band result in greater nonlinear distortions, limiting throughput.
In \cite{10117379}, 144 coherent channels were transmitted with wavelengths ranging from 1283~nm to 1334~nm, achieving over 40~Tb/s over 45~km.
A considerable performance reduction of up to 6~dB for higher signal launch power was observed in the signal-to-noise ratio (SNR) around zero dispersion wavelengths due to fibre nonlinearity. 

Digital nonlinearity compensation techniques have been successfully implemented for coherent systems operating in the C-band, ranging from Volterra filters~\cite{Saavedra:19}, machine learning algorithms~\cite{Aparecido21}, and digital backpropagation (DBP)~\cite{Ip:08}.  In the DBP algorithm, the signal is digitally back-propagated over a virtual link, where the nonlinear Schr\"{o}dinger equation (NLSE) is solved in reversed steps and with inverted parameters, allowing full compensation of the linear and nonlinear deterministic effects.  This technique, although very powerful, can present a high computational complexity as it requires two fast Fourier transform operations per step and multiple steps per span~\cite{Liga:14}.  However,  the low chromatic dispersion in the O-band has the advantage of allowing a significant reduction in complexity of the DBP digital signal processing (DSP), since a smaller number of steps is required to solve the NLSE with sufficient accuracy to mitigate the nonlinearities. 

In this paper we investigate the application of DBP for nonlinearity compensation in the transmission of dual-polarisation QAM signals with coherent detection in the O-band.  Our experiments with 25- and 50-GBd signal transmission over a 2-span, 151-km long Corning\textsuperscript{{\textregistered}} SMF-28\textsuperscript{{\textregistered}} ULL G.654.C-compliant fibre link with chromatic dispersion of up to 2.5~ps/(nm$\cdot$km) show that one-step DBP provides significant self-phase modulation (SPM) compensation.

\section{Experimental Setup}

\noindent Figure~\ref{fig:setup} shows the experimental transmission setup. Light from a \textless10~kHz linewidth ECL was first split by a PM 3-dB coupler and independently modulated by 23-GHz bandwidth IQ-MZMs to generate 25-GBd or 50-GBd 256QAM signals, shaped with a 1\%-roll-off root-raised cosine filter. The QAM signals were generated offline, applying linear digital pre-distortion~\cite{10132876}, and the modulators were driven by a 92-GS/s arbitrary waveform generator with an ENOB of 5-bits. The signals were re-combined with a polarisation beam combiner to form a dual-polarisation QAM signal. A BDFA was used as a booster amplifier, followed by a VOA to control the launched optical power (LOP). 

The BDFA comprised a 200-metre bismuth-doped fibre (BDF), backward-pumped by a 1150~nm beam, offering \textgreater20-dB gain with a 5-dB NF with a design similar to that presented in~\cite{10313970}. 
The gain and NF versus wavelength for an input power of 0~dBm are shown in Fig.~\ref{fig:setup}.
The signals were transmitted over a 151-km link, comprising 2 spans of 75.5-km of SMF-28 ULL fibre with an attenuation of 0.283~dB/km at 1310~nm and a mid-link BDFA. Although SMF-28 ULL fibre may have a cutoff wavelength up to 1520~nm (note that G.654.E specification allows up to 1530~nm), negligible penalty due to multipath interference is expected in the O-band \cite{10117198}.

After transmission, the fibre attenuation was compensated by a third BDFA. A second VOA controlled the power into the receiver, followed by an optical bandpass filter to remove the out-of-band amplified spontaneous emission generated by the BDFAs.
Power monitors were used before the transmission block, between the two spans, and at the receiver to assist in controlling the launched and received power, the latter being maintained at~2~dBm.

The coherent receiver consisted of a 90\degree-hybrid, four 70-GHz balanced photodetectors, and a 33\nobreakdash-GHz digital sampling oscilloscope with 8-bit ADCs.
The lower oscilloscope bandwidth (compared to the photodetectors) was due to practical limitations, as a single oscilloscope was used to simultaneously capture all four tributaries.
A \textless10~kHz linewidth O-band ECL was used as a local oscillator (LO).
Following detection, pilot-based DSP was used with a pilot sequence length of $2^{10}$ and a pilot insertion rate of 1/32 ~\cite{Wakayama:21}. EDC or DBP was subsequently applied, with EDC compensating for residual chromatic dispersion, and DBP addressing both dispersion and nonlinear effects.
Transmission performance was evaluated at three wavelengths: 1290~nm, 1310~nm, and 1330~nm.
An initial launch power versus SNR sweep was performed using EDC to determine the optimal launch power. 
DBP was then applied at this power, sweeping the dispersion and nonlinear coefficient ($\gamma_{\text{DBP}}$) used in the algorithm to determine the values that maximise the SNR gain relative to EDC.
The optimal chromatic dispersion values at 1290~nm, 1310~nm, and 1330~nm employed in the DBP were found to be \textminus2.5, 0.01, and 2.2~ps/(nm·km), respectively.
The DBP was implemented following the Wiener-Hammerstein (WH) model~\cite{5467218}, illustrated in Fig.~\ref{fig:WH}, 
where a fraction $\kappa$ of the link accumulated dispersion is first compensated, followed by nonlinearity compensation (a signal-power-dependent phase shift), and finally the remaining $(1-\kappa)$ portion of the dispersion is compensated in the final stage.
This single step combines the individual lengths and LOPs to account for both spans
The SNR was defined as $\mathrm{SNR} = {\mathbb{E}[|X|^2]}/{\mathbb{E}[|X-Y|^2]}$, where $X$ and $Y$ are the transmitted and received signals, respectively.


\section{O-band Transmission Results}
\begin{figure*}[t]
    \vspace{-1em}
    \centering
    \resizebox{0.95\textwidth}{!}{\input{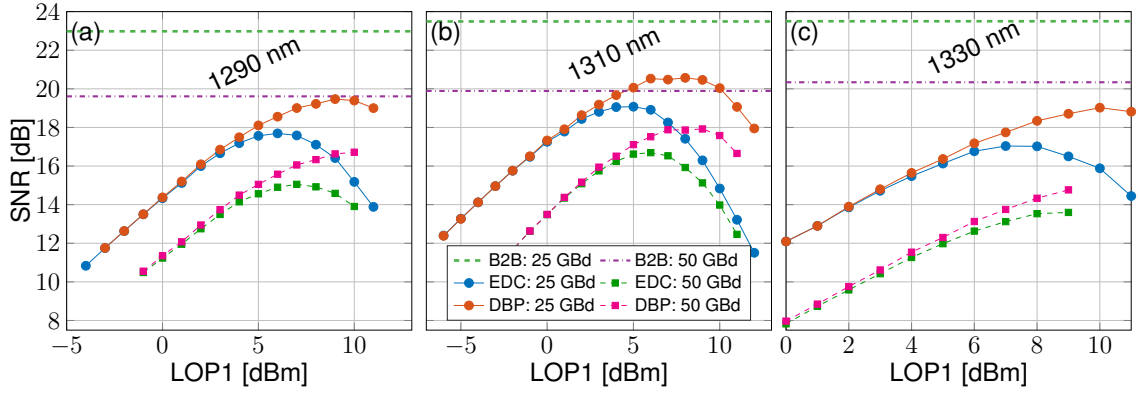}}
    \caption{ SNR versus first span launch optical power (LOP1) after 256QAM signals transmission over two spans, comparing EDC and DBP with single step at 25~GBd and 50~GBd for (a) 1290~nm, (b) 1310~nm, and (c) 1330~nm.}
    \label{fig:256qam}  
\end{figure*}

\noindent The back-to-back (B2B) SNR values were measured for 25~GBd and 50~GBd at 1290~nm, 1310~nm, and 1330~nm for the PDM-256QAM modulation format, and the results are shown in the table in Fig~\ref {fig:setup}. Each  B2B SNR value is the average over 100 recorded traces.
The SNR after transmission was measured over a range of launched optical powers after EDC and DBP for PDM-256QAM signals at both 25~GBd and 50~GBd, and the results are shown in Fig.~\ref{fig:256qam}, for the 1290~nm, 1310~nm, and 1330~nm wavelengths, respectively.
Note that, since the BDFA gain was lower than the loss of the first fibre span, the launch power into the second span was lower than that into the first span, with both sets of powers shown in Table \ref{table:LOP}.

\begin{table}[H]
\centering
\caption{Launch powers  into the first span (LOP1) and the second span (LOP2) at the three different wavelengths.}
\scriptsize 
\resizebox{0.9\linewidth}{!}{%
\begin{tabular}{|c|c|c|c|}
\hline
               & 1290 {[}nm{]}  & 1310 {[}nm{]}  & 1330 {[}nm{]}  \\ \hline
LOP1 {[}dBm{]} & LOP2 {[}dBm{]} & LOP2 {[}dBm{]} & LOP2 {[}dBm{]} \\ \hline
-9             & -1.94          & -0.33          & -2.75          \\ \hline
-6             & -1.73          & 0              & -2.56          \\ \hline
-3             & -1.3           & 0.36           & -2.3           \\ \hline
0              & -0.8           & 0.96           & -1.94          \\ \hline
3              & 0.2            & 1.93           & -1.42          \\ \hline
6              & 1.6            & 3.3            & -0.6           \\ \hline
9              & 3.42           & 5              & 0.7            \\ \hline
\end{tabular}
}
\label{table:LOP}
\end{table}

Transmission performance was assessed by computing the SNR as the average over 50 recorded traces.
At 1290 nm, the SNR gain from DBP was approximately 1.8~dB at 25~GBd, measured at the optimal LOP, which resulted in the highest SNR. 
However, at 50~GBd, the system was limited to a greater extent by the transceiver noise, and the DBP gain was reduced to around 1.6 dB.
For 1330~nm, an SNR gain of 2.0~dB was observed for the 25~GBd case, while for 50~GBd  the available LOP was insufficient to assess the SNR at the optimal power; nonetheless, at 9~dBm, a gain of approximately 1.1~dB was measured by applying DBP.
At 1310~nm, where the chromatic dispersion is near zero, a gain of 1.6~dB was observed for 25~GBd symbol rate, while for 50~GBd, 1.2~dB SNR gain was demonstrated. 
In terms of achievable information rate (AIR)\cite{8240991}, calculated from the generalised mutual information, increases of 11.3\%, 7.6\%, and 7.7\% were achieved for 50~GBd signals at 1290~nm, 1310~nm, and 1330~nm, respectively, with AIRs improving from 440~Gb/s, 494~Gb/s, and 396~Gb/s to 496~Gb/s, 535~Gb/s, and 430~Gb/s, respectively.

\begin{figure}[H]
  \centering
  \begin{minipage}{\linewidth}
    \centering
    \includegraphics[width=0.9\textwidth]{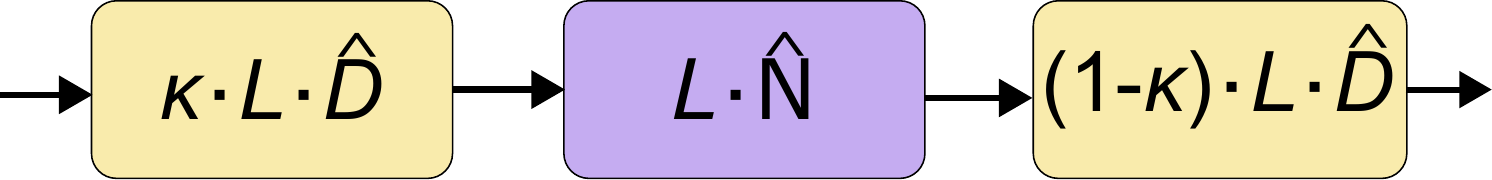}
    \vspace{-1em}
    
    \resizebox{\textwidth}{!}{
%
%
\definecolor{mycolor1}{rgb}{0.00000,0.44700,0.74100}%
\definecolor{mycolor2}{rgb}{0.85000,0.32500,0.09800}%
\definecolor{mycolor3}{rgb}{0.92900,0.69400,0.12500}%
\begin{tikzpicture}

\begin{axis}[%
  width=6.999in,
  height=5in,
  at={(1.174in,0.732in)},
  scale only axis,
  xmin=-0.01,
  xmax=1.01,
  xlabel style={font=\Huge, color=white!15!black, yshift=-10pt},
  xlabel={$\text{Wiener-Hammerstein split (}\kappa\text{)}$},
  ymin=1,
  ymax=2.2,
   xtick={0,0.2,0.4,0.6,0.8,1},
  ylabel style={font=\Huge, color=white!15!black , yshift=15pt},
  ylabel={SNR gain [dB]},
  ytick={1, 1.2, 1.4, 1.6, 1.8, 2, 2.2},
  tick label style={font=\Huge},
  legend style={font=\Huge, legend cell align=left, align=left, draw=white!15!black},
  axis background/.style={fill=white},
  axis x line*=bottom,
  axis y line*=left,
  xmajorgrids,
  ymajorgrids,
  legend style={legend cell align=left, align=left, draw=white!15!black, font=\huge, legend columns=2, at={(0.98,0.26)}, anchor=south east}
]
\addplot [color=mycolor1, line width=3.5pt, mark=o, mark size=4pt, mark options={solid, mycolor1}]
  table[row sep=crcr]{%
0	1.81506194566673\\
0.2	1.81412903468948\\
0.4	1.81412903468948\\
0.6	1.81412903468948\\
0.8	1.81412903468948\\
1	1.81412903468948\\
};
\addlegendentry{1290 nm - 25 GBd}

\addplot [color=mycolor1, line width=3.5pt, dashed, mark=square, mark size=4pt, mark options={solid, mycolor1}]
  table[row sep=crcr]{%
0	1.59049790497354\\
0.2	1.59049790497354\\
0.4	1.59049790497354\\
0.6	1.59049790497354\\
0.8	1.59049790497354\\
1	1.59049790497354\\
};
\addlegendentry{1290 nm - 50 GBd}

\addplot [color=mycolor2, line width=3.5pt, mark=o, mark size=4pt, mark options={solid, mycolor2}]
  table[row sep=crcr]{%
0	1.5740113587846\\
0.2	1.57388366325182\\
0.4	1.57388366325182\\
0.6	1.57388366325182\\
0.8	1.57388366325182\\
1	1.57388366325182\\
};
\addlegendentry{1310 nm - 25 GBd}

\addplot [color=mycolor2, line width=3.5pt, dashed, mark=square, mark size=4pt, mark options={solid, mycolor2}]
  table[row sep=crcr]{%
0	1.23909891781804\\
0.2	1.23855028297125\\
0.4	1.23855028297125\\
0.6	1.23855028297125\\
0.8	1.23855028297125\\
1	1.23855028297125\\
};
\addlegendentry{1310 nm - 50 GBd}

\addplot [color=mycolor3, line width=3.5pt, mark=o, mark size=4pt, mark options={solid, mycolor3}]
  table[row sep=crcr]{%
0	2.02877293260295\\
0.2	2.02924798507726\\
0.4	2.02924798507726\\
0.6	2.02924798507726\\
0.8	2.02924798507726\\
1	2.02924798507726\\
};
\addlegendentry{1330 nm - 25 GBd}

\addplot [color=mycolor3, line width=3.5pt, dashed, mark=square, mark size=4pt, mark options={solid, mycolor3}]
  table[row sep=crcr]{%
0	1.07727446130474\\
0.2	1.08211652837759\\
0.4	1.08211652837759\\
0.6	1.08211652837759\\
0.8	1.08211652837759\\
1	1.08211652837759\\
};
\addlegendentry{1330 nm - 50 GBd}

\end{axis}

\begin{axis}[%
width=9.031in,
height=6.656in,
at={(0in,0in)},
scale only axis,
xmin=0,
xmax=1,
ymin=0,
ymax=1,
axis line style={draw=none},
ticks=none,
axis x line*=bottom,
axis y line*=left
]
\end{axis}
\end{tikzpicture}%
}

  \end{minipage}
  \caption{Top figure: Wiener–Hammerstein model diagram. Bottom figure: SNR gain from DBP over EDC as a function of the WH split at the optimal DBP LOP.}
  \label{fig:WH}
  \vspace{-0.05em}
\end{figure}
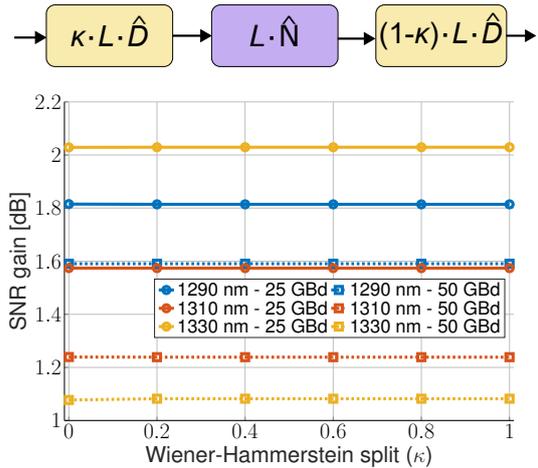

All nonlinearity-compensated signals shown in Fig.~\ref{fig:256qam} were measured using a WH split of $\kappa=0.5$. 
The resulting DBP SNR gain, relative to EDC, is presented in Fig.~\ref{fig:WH} for a range of WH split from 0 to 1. 
The variation in split ratio had a negligible impact on performance, indicating that dispersion can be fully compensated either before or after the nonlinearity compensation. 
This finding significantly simplifies the implementation, as it halves the number of FFT operations required.

In the case of wavelength division multiplexed transmission, additional interchannel nonlinear effects, such as cross-phase modulation and four-wave mixing, would lower the optimal launch powers per channel and hence would affect the SNR gains achievable with the use of single channel DBP. 
The low dispersion in the O-band leads to these interchannel effects being greater than in the C-band, and an assessment of this will be the subject of future experimental investigation.

\section{Conclusions}
\noindent To the best of our knowledge, this work represents one of the first demonstrations of digital backpropagation for nonlinearity mitigation in the O-band and near-zero dispersion regime. The low chromatic dispersion in the O-band increases the practical feasibility of DBP, as it allows the use of a single step. Low-complexity nonlinearity compensation based on a single-step DBP was demonstrated in PDM-256QAM signal transmission over a 151~km ULL G.654.C-compliant fibre link, leading to an SNR gain of up to 1.6~dB and a corresponding 11.3\% increase in the achievable information rate at 50~Gbaud.

\clearpage
\section{Acknowledgements}
\noindent
This work was supported by EPSRC Grant EP/R035342/1 Transforming Networks - building an intelligent optical infrastructure (TRANSNET), EP/W015714/1 Extremely Wideband Optical Fibre Communication Systems (EWOC), and EP/V007734/1 EPSRC Strategic equipment grant.

Romulo Aparecido is supported by a UCL Research Excellence Scholarship, and Polina Bayvel, by a Royal Society Research Professorship.

\bibliography{references}
\vspace{-4mm}

\end{document}